# Deterministic and stochastic properties of self-similar Rayleigh-Taylor mixing induced by space-varying acceleration

Arun Pandian (1), Snezhana I. Abarzhi (2)*

(1) Carnegie Mellon University, USA; (2) The University of Western Australia, AUS

* corresponding author email: snezhana.abarzhi@gmail.com

Rayleigh-Taylor interfacial mixing has critical importance in a broad range of processes in nature and technology. In most instances Rayleigh-Taylor dynamics is induced by variable acceleration, whereas the bulk of existing studies is focused on the cases of constant and impulsive accelerations referred respectively as classical Rayleigh-Taylor and classical Richtmyer-Meshkov dynamics. In this work we consider Rayleigh-Taylor mixing induced by variable acceleration with power-law dependence on the spatial coordinate in the acceleration direction. We apply group theory and momentum model to find deterministic asymptotic solutions for self-similar RT mixing. We further augment momentum model with a stochastic process to study numerically the effect of fluctuations on statistical properties of self-similar mixing in a broad parameter regime. We reveal that self-similar mixing can be Rayleigh-Taylor-type and Richtmyer-Meshkov type depending on the acceleration exponent. We further find the value of critical exponent separating Rayleigh-Taylor-type mixing and Richtmyer-Meshkov-type mixing, and identify invariant quantities characterizing Rayleigh-Taylor-type mixing and Richtmyer-Meshkov-type mixing.

1. Introduction
2. Modeling of deterministic and stochastic properties of RT mixing with space-varying acceleration
    2.1 Governing equations and theoretical modeling approaches
    2.2 Momentum model
    2.3 Stochastic model
3. Deterministic dynamics of RT mixing with space-varying acceleration
    3.1 Solution by method of Lie groups
    3.2 RT-type mixing
    3.3 RM-type mixing
    3.4 RT-RM transition
    3.5 Dimensionless invariant quantities in RT-type and RM-type mixing
4. Stochastic dynamics of RT mixing with space-varying acceleration
    4.1 Numerical model and computational set-up
    4.2 Parameter study
    4.3 RT-type mixing
    4.4 RM-type mixing
    4.5 Dimensionless invariant quantities in RT-type and RM-type mixing
5. Discussion
6. Acknowledgements
7. References
8. Tables
9. Figure captions
    Figures

**Section 1 – Introduction**

Rayleigh-Taylor instability (RTI) develops when fluids of different densities are accelerated against their density gradient, and leads to intense interfacial Rayleigh-Taylor (RT) mixing of the fluids [1-3]. RTI and RT mixing have critical importance for a broad range of processes in fluids, plasmas and materials, in both natural and in artificial environments, in high and low energy density regimes, and at various scales, from astrophysical to molecular [3,4]. Examples include inertial confinement fusion, material transformation under impact, explosion of supernova, fossil fuel extraction, and many other processes [4-8]. Even RT are often driven by variable acceleration in realistic environments, most existing work on RT flows study the cases of sustained and impulsive accelerations, referring to the former as classical Rayleigh-Taylor instability and to the latter as to classical Richtmyer-Meshkov (RM) instability [1,2,9,10].

In this work we focus on RT mixing with acceleration varying as power-law with space coordinate in the acceleration direction, and perform a detailed analytical and numerical study of deterministic and stochastic properties of RT dynamics within the frame of momentum model [11-18]. We reveal that there are two sub-regimes of RT-mixing that's set by the acceleration exponent: RT-type mixing and RM-type mixing. We find the critical exponent separating RT and RM type mixing and identify invariant quantities characterizing each sub-regime [16-18].

Rayleigh-Taylor instability starts to develop when the interface between two fluids with different densities is slightly perturbed near its equilibrium state [1,2]. The interface perturbations grow with time; a large-scale structure of bubble and spikes develops at the interface, with the light (heavy) fluid penetrating the heavy (light) fluid in bubbles (spikes) [3,19]. The large-scale dynamics is usually ordered and is sensitive to deterministic (initial) conditions [3,19,20]. It is characterized by the contribution of two relatively independent macroscopic scales - the amplitude in the direction of the acceleration and the spatial period in the normal plane [3,19]. Small-scale vortical structures are produced by shear at the interface [19-22]. Intensive interfacial mixing develops with time. In RT mixing, the amplitude is believed to grow self-similarly, whereas the dynamics of the spatial period depends on spectral properties and on symmetry of deterministic (initial) conditions [1-3,12,19-24].

There are the challenges in studying RT flows: the demanding requirements for the flow control and diagnostics in experiments [20-22,25-29], the necessities to accurately capture interfaces and dissipation processes in simulations [23,24,30-33], and the needs to account for non-local and singular character of the interface evolution in theory [3,19,34-38]. A remarkable success has been recently achieved in understanding of fundamentals of RT mixing [12,20]. Particularly, by employing group theory and momentum model, theoretical analysis [3,11,12,20] revealed that symmetries, invariants, and scaling and spectral properties of RT mixing differ substantially from those of canonical turbulence [34,35]. Theory found that RT mixing with constant acceleration may keep order, since it has stronger correlations, weaker fluctuations and stronger sensitivity to deterministic conditions when compared to isotropic homogeneous turbulence. The theory explained a broad range of experiments in fluids and plasmas, which observed that even at high Reynolds numbers RT mixing may be dominated by coherent structures [20-22,26-29], and simulations, which noted departures of RT dynamics from traditional turbulent scenario [13,23,24,30-33]. Only limited information is currently available on RT dynamics with variable acceleration suggesting a need in a systematic analysis [18]. Recently, a theoretical study has been conducted of RTI and RT mixing with time-varying acceleration [16-18]. The focus of the present work is on space-varying acceleration with a power-law dependence on spatial coordinate in the acceleration direction. On the side of fundamentals, power-law functions are important to study because they lead to new invariant and scaling properties of the dynamics [35]. In applications, power-law functions can be used to adjust the acceleration's time-dependence in realistic environments and ensure practicality of our results [5-8].

The other important aspect is the effect of fluctuations on properties of RT mixing [14,15,17,37]. Appearance of fluctuation in RT flows is usually associated with shear-driven interfacial vortical structures and with broad band initial perturbations [20-33,36]. While it is commonly believed that the former may produce small scale irregularities, the latter may enhance the interactions of large scales, and that both may transfer RT flow to a turbulent-like state, the properties of fluctuations in RT mixing flow is still a puzzle [3,12,17,18,20-33,36,37]. We need to better understand: What is the source of fluctuations in RT flows? Is RT mixing chaotic and sensitive to deterministic (initial) conditions? Is it stochastic and independent of deterministic (initial) conditions? Since in experiments and simulations statistical properties of RT dynamics are a challenge to quantify, there is a strong need in theory [3,4,19-21].

In this work we study deterministic and stochastic properties of RT mixing with a power-law space-varying acceleration within the frames of group theory and momentum model [11-18]. This approach has worked remarkably well for RTI and RT mixing with constant and time-varying acceleration [3,11-14,19,20]. The momentum model balances, per unit mass, the rates of gain and loss of momentum [3,11,12]. It is represented by the nonlinear inhomogeneous differential equations with the same symmetries and scaling transformations, as the equations for the conservation laws of mass, momentum and energy [18]. To account for the effect of fluctuations we augment the momentum model with multiplicative noise and analyze the corresponding set of nonlinear stochastic differential equations [14,15,17]. For a broad range of the model parameters, analytical solutions are identified for self-similar mixing dynamics. By applying stochastic model [14,15,17], solutions' statistical properties are investigated. We find that depending on the acceleration and each sub-regime has its own invariance quantity [18].

**Section 2 – Modeling of deterministic and stochastic properties of Rayleigh-Taylor mixing with space-varying acceleration**

Sub-section 2.1 - Governing equations and theoretical modeling approaches

For ideal fluids, dynamics of the Rayleigh-Taylor instability and RT mixing are governed by the conservation of mass, momentum and energy:

$$\partial \rho/\partial t + \partial \rho v_i/\partial x_i = 0, \partial \rho v_i/\partial t + \sum_{j=1}^{3} \partial \rho v_i v_j/\partial x_i + \partial P/\partial x_i = 0, \partial E/\partial t + \partial (E+P) v_i/\partial x_i = 0 \quad (1.1)$$

with spatial coordinates and time $(x_i, t) = (x, y, z, t)$, with fields of density, velocity, pressure and energy $(\rho, \mathbf{v}, P, E)$, with $E = \rho(e + \mathbf{v}^2/2)$ and $e$ being specific internal energy [3,18,19]. Upon introducing a local scalar function $\theta(x, y, z, t)$ with $\theta = 0$ at the interface, and with heavy (light) fluid located at $\theta > 0$ ($\theta < 0$) and marked with sub-script(s) $h(l)$, and upon presuming the zero mass flux across the interface and the absence of external sources, boundary conditions at the interface and at outside boundaries of the domain are

$$[\mathbf{v} \cdot \mathbf{n}] = 0, [P] = 0, [\mathbf{v} \cdot \boldsymbol{\tau}] = arbitrary, [W] = arbitrary, \quad \mathbf{v}_h \big|_{z \to -\infty} = 0, \mathbf{v}_l \big|_{z \to +\infty} = 0 \quad (1.2)$$

where $[...]$ denotes the jump of functions across the interface, $\mathbf{n}(\boldsymbol{\tau})$ is the unit normal (tangential) vector of the interface, $\mathbf{n} = \nabla \theta / |\nabla \theta| \, (\mathbf{n} \cdot \boldsymbol{\tau} = 0)$ and $W = e + P/\rho$ is the specific enthalpy [3,18,19]. Initial conditions include initial perturbations of the flow fields in the bulk and at the interface. The flow is periodic in the plane $(x, y)$, and is subject to acceleration (gravity) $\mathbf{g}$ directed from the heavy to the light fluid along the $z$-axis. Initial conditions are initial perturbations of the flow fields and the interface, including the spatial period (wavelength) $\lambda$ of the initial perturbation and the initial growth-rate is $v_0$ [3,18,19].

Even for ideal incompressible fluids, with $\dot{e} = 0$ and $\nabla e = 0$, theoretical problem Eq.(1) is intellectually rich: One has to solve a system of nonlinear partial differential equations in four-dimensional space-time, solve the boundary value problems for a sub-set of nonlinear partial differential equations at a nonlinear freely evolving interface and at the outside boundaries with account for secondary instabilities and singularities, and also solve the ill-posed initial value problem [3,13,18,19]. Nevertheless, with this complexity, dynamics RTI and RT mixing have remarkable features of universality and order and can thus be considered from the first principles by means of group theory [3,19]. For linear and nonlinear RTI, this approach applies theory of discrete groups to solve the boundary value and initial value problems Eq.(1) [3,19,35]. For RT mixing, group theory is implemented in momentum model, whose equations have the same symmetries and scaling transformations as the conservation laws [3,11,12,20].

Momentum model [3,11-13,18,20] considers balance, per unit mass, of the rates of gain and loss of specific momentum. For constant acceleration, momentum model finds the invariance of the rate of momentum loss, and identifies scaling and spectral properties of RT [3,12,20]. Principal results of momentum model - ordered character of RT mixing at high Reynolds numbers, and existence of a quasi-turbulent state with a short dynamic range for low-to-moderate Reynolds numbers - explained a broad set of existing experiments and simulations [3,12,20-33]. Momentum model has been applied to RT mixing with time-varying acceleration [16-18].

In this work, we apply the momentum model to the case of space-varying acceleration, and find deterministic and stochastic properties of RT mixing with space-varying acceleration, which, to our knowledge, have not been discussed in details before [18].

Sub-section 2.2 - Momentum model

Dynamics of a parcel of fluid undergoing RT mixing is governed by a balance per unit mass of the rates of momentum gain, $\mu$, and the rate of momentum loss, $\tilde{\mu}$, as

$$\dot{h} = v, \dot{v} = \tilde{\mu} - \mu \quad (2.1)$$

Here $h$ is the vertical length scale along the acceleration $\mathbf{g}$, $v$ is the corresponding velocity, $\tilde{\mu}$ and $\mu$ are the magnitudes of the rates of gain and loss of specific momentum in the vertical direction, and dot marks a (partial) time derivative [3,11,12,16-18,20]. The rates of gain $\tilde{\mu}$ and loss $\mu$ of specific momentum are associated with the rates of gain and loss of specific energy, $\tilde{\varepsilon}$ and $\varepsilon$, as $\tilde{\mu} = \tilde{\varepsilon}/v$ and $\mu = \varepsilon/v$. The rate of gain of specific energy is $\tilde{\varepsilon} = fgv$, where $g = |\mathbf{g}|$, and factor $f$ depends on the density ratio. Hereafter we re-scale $gf \to g$. We set $\varepsilon = Cv^3/L$, where $L$ is the characteristic length scale for energy dissipation, and drag $C \in (0,+\infty)$ is a model parameter [3,11,12,16-18,20]. This leads to $\tilde{\mu} = g$ and $\mu = Cv^2/L$

$$\dot{h} = v, \dot{v} = g - Cv^2/L \quad (2.2)$$

with $h = h(t)$, $v = v(t)$ and $h, v > 0$ for $t > t_0 > 0$.

Momentum model equations can also be presented for $v = v(h)$ as

$$v\,v' = g - Cv^2/L \quad (2.3)$$

with $v' = (dv/dh)$ and $(dv/dh) = (dh/dt)(dv/dt)$ and with $v > 0$ for $h > h_0 > 0$.

We consider accelerations with power-law dependence on the spatial coordinate in the acceleration direction, $g = g(h) = Gh^m$ with exponent $m$ and pre-factor $G > 0$ with dimensions $\dim m = 1$ and $\dim G = \mathrm{m}^{1-m}/\mathrm{s}^2$ [18].

The length scale $L$ for energy dissipation can be horizontal, $L \sim \lambda$ (the spatial period, or the wavelength), or vertical, $L \sim h$ (the amplitude), or a combination of scales, $L \sim L(\lambda, h)$ [3,11,12,16-18]. The case $L \sim \lambda \sim const$ corresponds to linear and nonlinear dynamics, and the case $L \sim h$ corresponds to the mixing dynamics [3,11,12,16-18]. Here we analyze asymptotic solutions for self-similar mixing with $L \sim h$, and we study the effect of fluctuations self-similar mixing, Figure 1 – Figure 3, Table 1 – Table 3.

Sub-section 2.2 - Stochastic model

To study the effect of fluctuations on RT mixing with space-varying acceleration, we augment the momentum model with the multiplicative noise, find numerical solutions for the system of nonlinear stochastic differential equations and investigate properties of the solutions.

We present the momentum model in a differential form

$$dh = v\,dt,\; dv = d\widetilde{M} - dM \quad (2.4)$$

assuming for differentials $d\widetilde{M} = g\,dt$ and $dM = C(v^2/h)g\,dt$, and considering $C(t)$ as a stochastic process [14,15,17,18].

The stochastic process is characterized by a stationary probability density function $p(C)$, and a time-scale $\tau_c$ describing how quickly the process $C(t)$ approaches $p(C)$. For deterministic dynamics the probability density function is delta-function, $p(C) = \delta(C - C_{max})$. For stochastic dynamics we consider the stationary probability density function $p(C)$ with log-normal distribution. The log-normal distribution is asymmetric, $\langle C \rangle \neq C_{max}$, with values $\langle C \rangle = C_0 e^{-\sigma^2/2}$ and $C_{max} = C_0 e^{-\sigma^2}$, where $\langle C \rangle$ is the mean of $C$, $C_{max}$ is the mode corresponding to the highest probability value of $p(C)$, and $\sigma$ is standard deviation. With $dW$ being the standard Wiener process, the time-evolution of stochastic process $C(t)$ obeys the equation [14,15,17,18]

$$dC = -C\left(ln(C/\langle C\rangle) - \sigma^2/2\right)dt/\tau_c + \sigma C\sqrt{2/\tau_c}\,dW \quad (2.5)$$

**Section 3 - Deterministic dynamics of self-similar mixing with space-varying acceleration**
Sub-section 3.1 - Solution by method of Lie groups

Momentum model has the same symmetries and scaling transformations as conservation laws, and captures, within a pre-factor, asymptotic behaviors of the dynamics [3,11,12,18-20].

In the mixing regime with $L = h$, the system is reduced to the first-order nonlinear non-homogeneous differential equation for $v = v(h)$ with $v' = (dv/dh)$ and $h, v > 0$

$$v' = Gh^m/v - Cv/h \quad (5.1)$$

It is also reduced to the second-order nonlinear non-homogeneous ordinary differential equation for $h = h(t), v = v(t)$, with $h, v, t > 0$.

$$\ddot{h} - C\dot{h}^2/h - Gh^m = 0 \quad (5.2)$$

To find the solutions for the equations, we apply the method of Lie groups [16-18].

Sub-section 3.2 - RT-type mixing

To find solution for the first-order space-dependent nonlinear non-homogeneous differential equation, we employ a one-parameter scaling transformation from $(v,h)$ to $(v_1,h_1)$ with parameter $\varepsilon$ and constant $\alpha$ as $h_1 = e^\varepsilon h$, $v_1 = e^{\alpha\varepsilon}v$. Under the transformation, the equation is modified to $e^{(1-\alpha)\varepsilon}\left[(dv_1/dh_1) - C(v_1/h_1)\right] - e^{(-m+\alpha)\varepsilon}h_1^m/v_1 = 0$; it is invariant under the scaling transformation at $\alpha = (1+m)/2$. This leads to $h_1 = e^\varepsilon h$, $v_1 = e^{\varepsilon(1+m)/2}v$ and identifies the equation invariant as $v^{2/(m+1)}/h = v_1^{2/(m+1)}/h_1$. Hence we find the solution $v_{RT} = N_{RT}h^{n_{RT}}$ for the non-homogeneous nonlinear differential equation with

$$v_{RT} = N_{RT}h^{n_{RT}}, n_{RT} = (m+1)/2, N_{RT} = (G/(m+C))^{1/2} \quad (6.1)$$

For this power-law solution the exponent $n_{RT}$ is set by the acceleration exponent $m$, and the pre-factor $N_{RT}$ depends on the acceleration exponent and strength $(m, G)$, and on drag $C$. The sub-script emphasizes that dynamics is driven by acceleration $g$ and is RT-type. Note that at $m = -1$ RT-type mixing is steady, $n_{RT} = 0$, Table 1.

Similarly we can find solution for the second-order time-dependent nonlinear non-homogeneous differential equation. We employ a one-parameter scaling transformation from $(h, t)$ to $(h_1, t_1)$ with parameter $\varepsilon$ and constant $\alpha$ as $t_1 = e^{\varepsilon} t, h_1 = e^{\alpha \varepsilon} h$. The equation gets the form $e^{(2-\alpha)\varepsilon} \left[ (d^2 h_1 / dt_1^2) - C((dh_1/dt_1)^2 / h_1) \right] - e^{-m\alpha\varepsilon} M h_1^m = 0$ ; it is invariant under the transformation at $\alpha = 2/(1-m)$, leading to $t_1 = e^{\varepsilon} t, h_1 = e^{2\varepsilon/(1-m)} h$ and the invariance of $h/t^{2/(1-m)} = h_1/t_1^{2/(1-m)}$. Hence we find the solution $h_{RT} = B_{RT} t^{b_{RT}}$ for the non-homogeneous nonlinear differential equation with

$$h_{RT} = B_{RT} t^{b_{RT}}, b_{RT} = 2/(1-m), B_{RT} = \left( (G/(m+C))^{1/2} / (2/(1-m)) \right)^{2/(1-m)} \quad (6.2)$$

and $b_{RT} = (1 - n_{RT})^{-1}$, $B_{RT} = (N_{RT}/b_{RT})^{b_{RT}}$. For this power-law solution, the exponent $b_{RT}$ is set by the acceleration exponent $m$, and the pre-factor $B_{RT}$ depends on the acceleration exponent and strength $(m, G)$ and on drag $C$. The sub-script emphasizes that dynamics is driven by the acceleration $g$ and is RT-type. In RT-type mixing the amplitude is an increasing function of time for $m < 1$. At $m = -1$ RT-type mixing is steady, $n_{RT} = 0$, Table 1.

Sub-section 3.3 - RM-type mixing

In addition to particular solution $v_{RT} = N_{RT} h^{n_{RT}}$, the first-order space-dependent non-homogeneous differential equation has a solution for the associated homogeneous equation $v' = -Cv/h$. This equation can be solved by applying the method of separation of variables, leading to $v = V_0 (h/H_0)^{-C}$ where $V_0, H_0$ are integration constants. We write the solution $v_{RM} = N_{RM} h^{n_{RM}}$ as

$$v_{RM} = N_{RM} h^{n_{RM}}, n_{RM} = -C, N_{RM} = V_0 H_0^C \quad (7.1)$$

For the power-law solution, the exponent $n_{RM}$ is set by drag $C$, and whereas the pre-factor $N_{RM}$ depends on the integration constants and on deterministic (initial) conditions. The sub-script emphasizes that dynamics is independent of acceleration $g$ and is RM-type, Table 1.

Similarly we can find solution for the associated homogeneous equation for the second-order nonlinear non-homogeneous differential equation $(d^2 h/dt^2) = -C((dh/dt)^2/h)$. It can be solved by applying the separation of variables $\left(h^{(1+C)} - H_0^{(1+C)}\right)/H_0^{(1+C)} = (1+C)V_0\left(t - \tilde{t}_0\right)$ with integration constants $V_0, H_0$ leading to asymptotic solution $h_{RM} = B_{RM} t^{b_{RM}}$ with

$$h_{RM} = B_{RM} t^{b_{RM}}, \, b_{RM} = (1+C)^{-1}, \, B_{RM} = H_0\left((1+C)(V_0/H_0)\right)^{1/(1+C)} \quad (7.2)$$

and $b_{RM} = (1 - n_{RM})^{-1}$, $B_{RM} = (N_{RM}/b_{RM})^{b_{RM}}$. For the power-law solution, the exponent $b_{RM}$ is set by drag $C$, and whereas the pre-factor $B_{RM}$ depends on the integration constants and on deterministic (initial) conditions. The sub-script emphasizes that dynamics is independent of the acceleration $g$ and is RM-type, Table 1.

Sub-section 3.4 - RT-RM transition

General solutions for the first-order space-dependent and second-order time-dependent non-homogeneous differential equations are some functions of the asymptotic RT and RM-type solutions, $v = v(v_{RT}(h), v_{RM}(h))$ and $h = h(h_{RT}(t), h_{RM}(t))$. RT and RM type solutions should be coupled since these equations are nonlinear. A remarkable property is that in a broad parameter regime RT-type and RM-type dynamics are effectively de-coupled due to their distinct symmetries. Particularly, RT-type dynamics is invariant with respect to a scaling transformation, whereas RM-type dynamics is invariant with respect to a point group associated with the arbitrariness of time origin [16-18].

By comparing exponents $n_{RT}$ and $n_{RM}$ as well as exponents $b_{RT}$ and $b_{RM}$, we find two sub-regimes of RT mixing with variable space-dependent acceleration. A critical exponent separating the sub-regimes is defined from the conditions $h^{n_{RT}} \sim h^{n_{RM}}$ with $h^{-C} \sim h^{2/(m+1)}$ and $t^{b_{RT}} \sim t^{b_{RM}}$ with $t^{(1+C)^{-1}} \sim t^{2/(m-1)}$ leading to

$$m_{cr} = -2C - 1 \quad (8)$$

with $m_{cr} \to -1$ for $C \to 0$ and $m_{cr} \to -\infty$ for $C \to \infty$.

For $m \in (m_{cr}, 1)$ RT-type mixing dominates the dynamics, and it has $\dot{v} > 0\,(<0)$ for $m > -1\,(<-1)$ and $\dot{v} = 0$ at $m = -1$. In RT-type mixing the terms $|\dot{v}|, |\tilde{\mu}|, |\mu| \sim h^m \sim t^{m \cdot b_{RT}}$ are asymptotically balanced with an algebraic imbalance of terms $|\tilde{\mu}| \neq |\mu|$. For $m \in (-\infty, m_{cr})$ RM-type mixing dominates the dynamics, and it has $\dot{v} < 0$. In RM-type mixing the terms $\dot{v}, \mu \sim h^m \sim t^{m \cdot b_{RT}}$ are asymptotically balanced with the asymptotic imbalance $|\tilde{\mu}| << |\dot{v}|, |\mu|$. At $m \sim m_{cr}$ a transition occurs from RT-type mixing to RM-type mixing.

Sub-section - 3.5 Dimensionless invariant quantities of RT-type and RM-type mixing

For RT-type mixing with $m > m_{crit}$, we rearrange terms in momentum model equation as

$$(\dot{v}/\tilde{\mu}) + (\mu/\tilde{\mu}) = (\dot{v}/g) + (Cv^2/gh) = 1 \quad (9.1)$$
$$(vv'/\tilde{\mu}) + (\mu/\tilde{\mu}) = (vv'/g) + (Cv^2/gh) = 1$$

In these equations, the first terms are $(\dot{v}/\tilde{\mu}) = (\dot{v}/g)$ and $(vv'/\tilde{\mu}) = (vv'/g)$ with $(\dot{v}/\tilde{\mu}) = (vv'/\tilde{\mu})$. By substituting the solution, we find that the first term is $(vv'/\tilde{\mu}) = (m+1)/(m - m_{cr})$. The second term is the ratio of the rates of loss and gain of specific momentum $\Pi = \mu/\tilde{\mu} = Cv^2/gh$, and it is $\Pi = (m_{cr} + 1)/(m_{cr} - m)$. Hence for RT-type mixing with $m_{cr} << m \leq 1$, value $\Pi \sim O(1)$. Furthermore, for reasonably large $C > 1$, value $\Pi \sim 1$ and value $|\dot{v}/g| << 1$ is small. This suggests that value $\Pi$ can be chosen as a dimensionless invariant quantity to characterize RT-type mixing for $m > m_{cr}$. To further justify the choice, we consider value $\Pi$ in sub-regime of RM-type mixing for $m \to m_{cr}^{\pm}$ and $m < m_{cr}$. We find that value $\Pi \sim h^{-(m_{cr} - m)/2} \to \infty$ in RM-type mixing sub-regime. Hence, value $\Pi$ is the proper dimensionless invariant quantity to characterize RT-type mixing with $m > m_{cr}$, Table 2.

For RM-type mixing, $m < m_{crit}$, we re-arrange terms in momentum model equation as

$$(\tilde{\mu}/\dot{v}) + (-\mu/\dot{v}) = (g/\dot{v}) + (-Cv^2/\dot{v}h) = 1 \quad (9.2)$$
$$(\tilde{\mu}/vv') + (-\mu/vv') = (g/vv') + (-Cv^2/vv'h) = 1$$

In these equations, the first terms are $\tilde{\mu}/\dot{v} = g/\dot{v}$ and $\tilde{\mu}/vv' = g/vv'$ with $\tilde{\mu}/vv' = \tilde{\mu}/\dot{v}$. The second terms are $\mu/\dot{v} = Cv^2/\dot{v}h$ and $\mu/vv' = Cv^2/vv'h$ with $\mu/\dot{v} = \mu/vv'$ and $\mu/\dot{v} = Cv^2/\dot{v}h$. By substituting the solution for $m < m_{crit}$, we find that term $\tilde{\mu}/\dot{v} = g/\dot{v}$ decays $|g/\dot{v}| \to 0$ as $|g/\dot{v}| \sim h^{-(m - m_{cr})}$, and term $\tilde{\Pi} = \mu/\dot{v} = Cv^2/\dot{v}h$ approaches unity, $\tilde{\Pi} \to 1$. This suggests that value $\tilde{\Pi}$ can be chosen as a dimensionless invariant quantity to characterize RM-type mixing with $m < m_{cr}$. To further justify this choice, we consider value $\tilde{\Pi}$ in RT-type mixing sub-regime for $m \to m_{cr}^{\pm}$ and $m > m_{cr}$. We find that for $m \to m_{cr}^{\pm}$ and $m > m_{cr}$, value $\tilde{\Pi} = (m_{cr}+1)/(m+1)$ is large, $|\tilde{\Pi}| \gg 1$. Hence, value $\tilde{\Pi}$ is the proper dimensionless invariant quantity to characterize RM-type mixing with $m < m_{cr}$, Table 2.

Section 4 - Stochastic dynamics

Sub-section 4.1 - Numerical model and computational set-up

To study the effect of fluctuations on the dynamics of RT mixing with space-varying acceleration, we numerically model the system of nonlinear stochastic differential equations with multiplicative noise and with power law acceleration $g(h) = Gh^m$. We modify the numerical solver [14,15,17,18] to reflect the initial conditions and equations for space-varying acceleration.

We consider a broad range of the acceleration exponents for space-varying acceleration with exponents from interval $m \in [-11,1)$ with $m = \{0.5, 0, -1, -3, -5, -9, -10, -11\}$. For the stochastic process $C(t)$ we set the time-scale $\tau_c = \tau$, where $\tau$ is time-scale of deterministic system, the mean value $\langle C \rangle = 3.6$, and the standard deviation $\sigma = \langle C \rangle / 2$. The mean value is chosen to provide the growth-rate coefficient as in typical observations of RT mixing with steady acceleration [14,15,17,18,23-33]. The initial conditions correspond to the asymptotic solution with value $C = 0$. The time-scale, the length-scale and the acceleration strength in the system are set to unity. In order to accurately and confidently identify statistical properties of the solutions, we use as default the number trajectories $N_v = 10^3$ and the span of scales of time of six decades, $t \in [10^0, 10^6]$. When averaged over the distribution $p(C)$, the value of critical exponent is $m_{cr} \approx -8.2$. Results presented below are averaged over $10^3$ trajectories unless specified otherwise. We run the numerical simulations on a supercomputer cluster.

Sub-section 4.2 – Parameter study

To ensure statistical confidence of our results, we study the effect of the model parameters, including the effects of the mean value $\langle C \rangle$, the number of trajectories $N_v$ and the span of scales $t \in [10^0, 10^n], n = 1, 2, ...6$.

To study the effect of the mean value $\langle C \rangle$, we vary $\langle C \rangle = \{1.0, 3.6, 5.0\}$ and compare the results with those at $\langle C \rangle = 3.6$ for RT-type mixing $m > m_{cr}$, and for RM-type mixing $m < m_{cr}$. We find that changes of value of $\langle C \rangle$ have little effect on RT-type mixing. RM-type mixing is more sensitive to variations of $\langle C \rangle$, due to the dependence of $m_{cr}$ on $C$: when $\langle C \rangle$ is small, larger fluctuations may appear in the modeled quantities; when $\langle C \rangle$ is large, the dynamics are smoother, since the large $\langle C \rangle$ acts as a fluctuation moderator.

To study the effect of the number of trajectories $N_v$, we consider the solutions for $10^0$, $10^1$, $10^2$ and $10^3$ trajectories. For small number of trajectories the process is statistically unconfident. For very large number of trajectories the computations are massive. To identify the statistically justified number of trajectories, and to isolate statistical unsteadiness, we use a steady state solution at $m = -1$. We observe the same general trend throughout all solutions with fewer trajectories leading to more noise in the measured values, and with standard deviations increasing for smaller number of trajectories.

We study the effect of the span of scales of time $t$ on our results. We set $m = -1$ for the statistically steady dynamics. The equations are simulated over six decades to see the behavior at different stages in the evolution. During the first-to-second decade, the dynamics are influenced by initial conditions. We may evaluate the mean past the second decade, and the statistical properties - past the third-to-fourth decade of time. The log-normal distribution of $C(t)/\langle C \rangle$ is identifiable at four decades.

Hence with $\langle C \rangle = 3.6$, with $N_v = 10^3$ trajectories and with the span of scales of six decades of time $t \in [10^0, 10^6]$, the results are statistically confident and the simulations are computationally efficient.

Sub-section 4.3 - RT-type mixing

To model RT-type mixing with space-varying acceleration, we consider the acceleration exponents $m = \{0.5, 0, -1, -3, -5\}$ with $m \in (m_{cr}, 1)$ and $m_{cr} \approx -8.2$ for $\langle C \rangle = 3.6$. Figure 1 presents the solution for a sample case $m = -3$. In Figure 1a, the amplitude $h$ increases with time, whereas the velocity $v$, the system acceleration $|\dot{v}|$ and the acceleration $g$ decrease with time, in agreement with our theoretical results. The exponents of the power-laws of $g$ and $\dot{v}$ are the same, whereas the pre-factor of $\dot{v}$ is smaller than that of $g$. Hence, properties of this asymptotic power-law solution are defined by the acceleration, and the dynamics is indeed driven by the acceleration and is RT-type, in agreement with our analytical results.

To determine if the numerically modeled solutions match the analytical solutions, and to evaluate sensitivity of the solutions to noise, we identify the exponent and the pre-factor of the numerical solutions and we scale the modeled values of the exponent and the pre-factor by their analytical values. For asymptotic solutions $h = Bt^b$ and $v = Bbt^{b-1}$. On the basis of simulations data the modeled values of exponent are defined as $b = (tv/h)$ and that of pre-factor as $B = (h/t^{(tv/h)})$. Figure 1b represents the values of the exponent and the pre-factor scaled with their corresponding analytical values. In these plots, an exact match between the numerical and analytical values corresponds to 1. As is seen from Figure 1b, the modeled exponent approaches the analytical value quickly, has smaller error throughout and fluctuates less than the pre-factor. The pre-factor, while approaching the analytical value, is more sensitive to noise. Note that in RT-type mixing the solution exponent depends only on the acceleration exponent, whereas the solution pre-factor depends on the acceleration parameters and on the fluctuating drag $C$. This suggests that the solution exponent is a reliable quantity to diagnose.

Sub-section 4.4 - RM-type mixing

To model RM-type mixing with variable space-dependent acceleration, we consider the acceleration exponents $m = \{-9, -10, -11\}$ with $m \in (-\infty, m_{cr})$ and $m_{cr} \approx -8.2$ for $\langle C \rangle = 3.6$. Figure 2 presents the solution for a sample case $m = -10$. In Figure 2a, the amplitude $h$ increases with time, whereas the velocity $v$, the solution's acceleration $|\dot{v}|$ and the system acceleration $g$ decrease with time, in agreement with our theoretical results. The exponents of the power-laws of $g$ and $\dot{v}$ are distinct, and the exponent of $g$ is smaller than that of $\dot{v}$. Hence, in agreement with our analytical results, properties of this asymptotic power-law solution are weakly influenced by the acceleration, and the dynamics is indeed RM-type.

Figure 2b presents the modeled exponent and pre-factor which we scale with their corresponding theoretical values. In RM-type mixing, both the modeled exponent and pre-factor approach their analytical values; the pre-factor retains the memory of the initial conditions. Note that in RM-type mixing the solution exponent and the solution pre-factor both depends on the (fluctuating) drag $C$, and the pre-factor also depends also on deterministic (initial) conditions. Hence, the solution exponent is a more reliable quantity to diagnose than a pre-factor.

Sub-section 4.5 - Dimensionless invariant quantities in RT-type and RM-type mixing

Figure 3 represent dimensionless quantities $\Pi = Cv^2/gh$ and $\tilde{\Pi} = -Cv^2/\dot{v}h$ in sample cases $m=-3, m>m_{cr}$ in RT-type mixing and $m=-10, m<m_{cr}$ in RM-type mixing with $m_{cr}=-2C-1$ and $m_{cr}=-8.2, \langle C \rangle = 3.6$. According to our analysis, in RT-type mixing the quantities are $\Pi \sim 1$ and $|\tilde{\Pi}| \gg 1$ with $\Pi = (m_{cr}+1)/(m_{cr}-m)$ and $\tilde{\Pi} = (m_{cr}+1)/(m+1)$, and in RM-type mixing the quantities are $\Pi \gg 1$ and $\tilde{\Pi} \sim 1$ with $\Pi \to \infty$ and $\tilde{\Pi} \to 1$. Numerical solutions agree with these analytical solutions. In Figure 3a at $m=-3, m>m_{cr}$, quantity $\Pi$ is $\Pi \sim 1$ and quantity $\tilde{\Pi}$ is $\tilde{\Pi} \gg 1$. In Figure 3b at $m=-10, m<m_{cr}$ quantity $\Pi$ is $\Pi \to \infty$ and quantity $\tilde{\Pi}$ is $\tilde{\Pi} \to 1$. Qualitatively similar results are obtained for other values of the acceleration's exponents.

Table 3 represents quantities $\Pi, \tilde{\Pi}$ averaged over the time interval $t \in [10^4, 10^5]$, as well as their analytical solutions for a broad range of acceleration exponents. Quantity $\Pi$ is $\Pi \sim O(1)$ in RT-type mixing with acceleration exponents $m \in (m_{cr}, 1)$; it start to diverge for $m \to m_{cr}$, and approaches $\Pi \to \infty$ in RM-type mixing with acceleration exponents $m \in (-\infty, m_{cr})$. Quantity $\tilde{\Pi}$ is $\tilde{\Pi} \to 1$ in RM-type mixing with acceleration exponents $m \in (-\infty, m_{cr})$, it start to grow for $m \to m_{cr}$, and it is $\tilde{\Pi} = (m_{cr}+1)/(m+1)$ with $|\tilde{\Pi}| \gg 1$ in RT-type mixing with $m \in (m_{cr}, 1)$.

Hence, RT-type mixing and RM-type mixing each has its own dimensionless invariant quantity. These are the quantities $\Pi = (Cv^2/gh)$ for $m>m_{cr}$ and $\tilde{\Pi} = (-Cv^2/\dot{v}h)$ for $m<m_{cr}$, each of which is the value comparable to unity in the appropriate type of mixing. At $m \to m_{cr}$ a transition occurs from RT to RM type mixing, Table 3.

**Section 5 - Discussion**

In this work, we analyzed self-similar Rayleigh-Taylor mixing induced by variable acceleration with power-law space-varying acceleration. We applied group theory and momentum model. Momentum model has the same symmetries and scaling transformations as the governing equations, and provides asymptotic solutions up to a constant. We performed a detailed study of deterministic and stochastic properties of RT dynamics. A broad range of parameters is investigated of the acceleration and of the stochastic process. To our knowledge, these results have not been obtained and discussed in details before. See Eqs.(1-9), Figures 1-3, Tables 1-3.

For self-similar RT mixing induced by space-varying power-law acceleration, there are two sub-regimes depending on the exponent of the acceleration power-law - Rayleigh-Taylor type mixing and Richtmyer-Meshkov type mixing. Self-similar RT-type mixing is described by a power-law solution, whose exponent is set by the acceleration exponent, and pre-factor - by the acceleration parameters and drag. Self-similar RM-type mixing is described by power-law solution independent of the acceleration parameters; its exponent is set by the drag, and pre-factor - by drag and by deterministic (initial) conditions. RT-RM transition occurs when the acceleration exponent approaches a critical value depending on the drag. While generally RT and RM type mixing are coupled, for acceleration exponents far from critical, the coupling is weak due to distinct symmetries of the solutions. We found explicit time and space dependence of solutions describing RT-type mixing and RM-type mixing and identified the range of acceleration exponents in which the amplitude is an increasing function of time, Eqs.(5-8). These theoretical benchmarks can be applied for a comparison with future experiments and simulations [18-33], and can serve to better understand RT-relevant processes in nature and technology [4-8].

We further studied the effect of fluctuations on RT-type mixing and RM-type mixing. We modeled the fluctuations by considering the drag as a stochastic process and found numerical solutions for the corresponding nonlinear stochastic differential equations. We performed a detailed study of parameters of the stochastic process, the span of scales and the number of trajectories on our numerical results, and identified the parameters' values at which the numerical results are statistically confident and simulations are computationally efficient. Our simulations achieve excellent agreement with the theory, find two sub-regimes of RT mixing with space-varying acceleration, and identify dimensionless invariant quantities characterizing each sub-regime. See Figure 1 – Figure 3 and Table 1 – Table 3. These benchmarks can serve for studies of statistical properties of RT mixing with space-varying acceleration in the future [18-33].

In self-similar RT-type mixing, the exponent of the power-law solution is set by the acceleration exponent and is a reliable diagnostic parameter insensitive to fluctuations; the pre-factor of the power-law solution depends on the drag and is more sensitive to fluctuations. These results are in excellent agreement with available observations [18-33]. They encourage the development of new methods for accurate diagnostics of statistical properties of RT-type mixing. See Figures 1-3, Tables 1-3.

In self-similar RM-type mixing the exponent of the power-law solution is set by the fluctuating drag, whereas the pre-factor depends on the drag and on its fluctuations and on deterministic (initial) conditions. This dynamics is universal since the growth of RM-type mixing has the same power-law exponent independent of the acceleration exponents. The dynamics is also non-universal since the pre-factor of the power-law solution depends also on deterministic (initial) conditions. Our results indicate that in RM-type mixing the exponent of the power-law is still a better diagnostic than a pre-factor, whereas the spread of exponent values is linked to the properties of fluctuations. These results indicate the need in further advancement of experimental and numerical methods for studying RT and RM mixing [18-33]. See Figures 1-3, Tables 1-3.

While RT-type mixing and RM-type mixing are well differentiated by exponents of their power-law solutions, there is a need in additional quantities to characterize RT-type and RM-type mixing. This is because accurate quantification of power-laws requires substantial span of scales [37], which are often a challenge to achieve in experiments and simulations, and because RT and RM mixing may co-exist when the acceleration exponent is close to the critical exponent. According to our results, RT-type mixing and RM-type mixing, each, has its own dimensionless invariant quantity. Its value is a number on the order of unity in the proper sub-regime, and diverges otherwise. Hence, by measuring these dimensional quantities, one can identify the sub-regime of RT mixing and, furthermore, can evaluate the magnitude of the drag. These results can serve to better understand the broad range of RT-relevant processes in nature and technology [5-10,18]. See Figures 1-3, Tables 1-3.

We consider space-varying acceleration for a broad range of acceleration parameters. In addition to critical exponent for transition from RT to RM-type mixing, there are other important particular cases. One such case is the exponent equals to 0. This case corresponds to classical RTI and RT mixing with constant acceleration. The other case is the exponent approaching negative infinity, which corresponds to classical RM mixing with impulsive acceleration. One more important case is the acceleration exponent equals to negative unity. This exponent value plays an important role in astrophysics, fusion, and nano-fabrication, since it is the limiting value for self-similar blast waves and impacts [18]. Moreover, at this exponent value the self-similar RT type mixing is free from the statistical unsteadiness since its dynamics is steady. See Eqs.(5-8).

To conclude, we analyzed the self-similar Rayleigh-Taylor mixing induced by power-law space-varying acceleration. Depending on the acceleration exponent, RT mixing can be RT-type or RM-type, with each sub-regime having its own characteristic solution and invariant quantity.

**Section 6 - Acknowledgements**

SIA thanks the National Science Foundation (USA), the University of Western Australia (AUS).

# Section 7 - References

**Section 8 - Tables**

Table 1: Self-similar mixing with variable acceleration $g = Gh^m$

| | RT-type mixing |
|---|---|
| | $m_{cr} < m < 1$ |
| $v = v(h)$ | $v_{RT} = N_{RT} h^{n_{RT}}$, $n_{RT} = (m+1)/2$, $N_{RT} = (G/(m+C))^{1/2}$ |
| $h = h(t)$ | $h_{RT} = B_{RT} t^{b_{RT}}$, $b_{RT} = 2/(1-m)$, $B_{RT} = \left((G/(m+C))^{1/2} / (2/(1-m))\right)^{2/(1-m)}$ |
| | RM-type mixing |
| | $m < m_{cr}$ |
| $v = v(h)$ | $v_{RM} = N_{RM} h^{n_{RM}}$, $n_{RM} = -C$, $N_{RM} = V_0 H_0^C$ |
| $h = h(t)$ | $h_{RM} = B_{RM} t^{b_{RM}}$, $b_{RM} = (1+C)^{-1}$, $B_{RM} = H_0 ((1+C)(V_0/H_0))^{1/(1+C)}$ |

Table 2: Characteristic quantities of self-similar mixing with variable acceleration $g = Gh^m$

| Regime | Exponents | $\Pi = Cv^2/gh$ | $\tilde{\Pi} = -Cv^2/\dot{v}h$ |
|---|---|---|---|
| RT-type | $m_{cr} \ll m < 1$ | $\Pi \sim O(1)$, $\Pi = \dfrac{m_{cr}+1}{m_{cr}-m}$ | $\lvert\tilde{\Pi}\rvert \gg 1$, $\tilde{\Pi} = \dfrac{m_{cr}+1}{m+1}$ |
| RM-type | $-\infty < m \ll m_{cr}$ | $\Pi \to \infty$, $\Pi \sim h^{-(m_{cr}-m)/2}$ | $\tilde{\Pi} \to 1$, $\tilde{\Pi} = \dfrac{m_{cr}+1}{m_{cr}+1}$ |

Table 3: Characteristic quantities of self-similar mixing with variable acceleration $g = Gh^m$ in the simulations averaged over time interval $[10^4, 10^5]$ (theory values are in parenthesis)

| $m$ | 0.5 | 0 | -1 | -3 | -5 | -9 | -10 | -11 |
|---|---|---|---|---|---|---|---|---|
| $\Pi$ | 0.826 (0.828) | 0.877 (0.878) | 0.999 (1) | 1.39 (1.38) | 2.29 (2.25) | 2.6x10³ (∞) | 5.8x10⁵ (∞) | 2.3x10⁷ (∞) |
| $\tilde{\Pi}$ | -4.78 (-4.8) | -7.22 (-7.2) | (±∞) | 3.57 (3.6) | 1.78 (1.8) | 1.00 (1) | 1.00 (1) | 1.00 (1) |

## Section 9 - Figure captions and Figures

Figure 1: Asymptotic power-law solutions, averaged over 1000 trajectories for Rayleigh-Taylor type mixing with acceleration's exponent (-3): (a) amplitude (red), velocity (blue) and solution' acceleration (green) as well as the system acceleration (black); (b) exponent (blue) and pre-factor (green) scaled with their analytical values. Power-law solution is clearly seen from log-log plot. The solution's exponent is a robust and reliable parameter to diagnose.

Figure 2: Asymptotic power-law solutions, averaged over 1000 trajectories for Richtmyer-Meshkov type mixing with acceleration's exponent (-10): (a) amplitude (red), velocity (blue) and solution' acceleration (green) as well as the system acceleration (black); (b) exponent (blue) and pre-factor (green) scaled with their analytical values. Power-law solution is clearly seen from log-log plot. The solution's exponent is a robust and reliable parameter to diagnose.

Figure 3: Characteristic quantities for: (a) Rayleigh-Taylor type mixing with acceleration's exponent (-3); (b) Richtmyer-Meshkov type mixing with acceleration's exponent (-10). The quantity is an invariant value on the order of 1 in the proper regime and diverges otherwise.

Figure 1

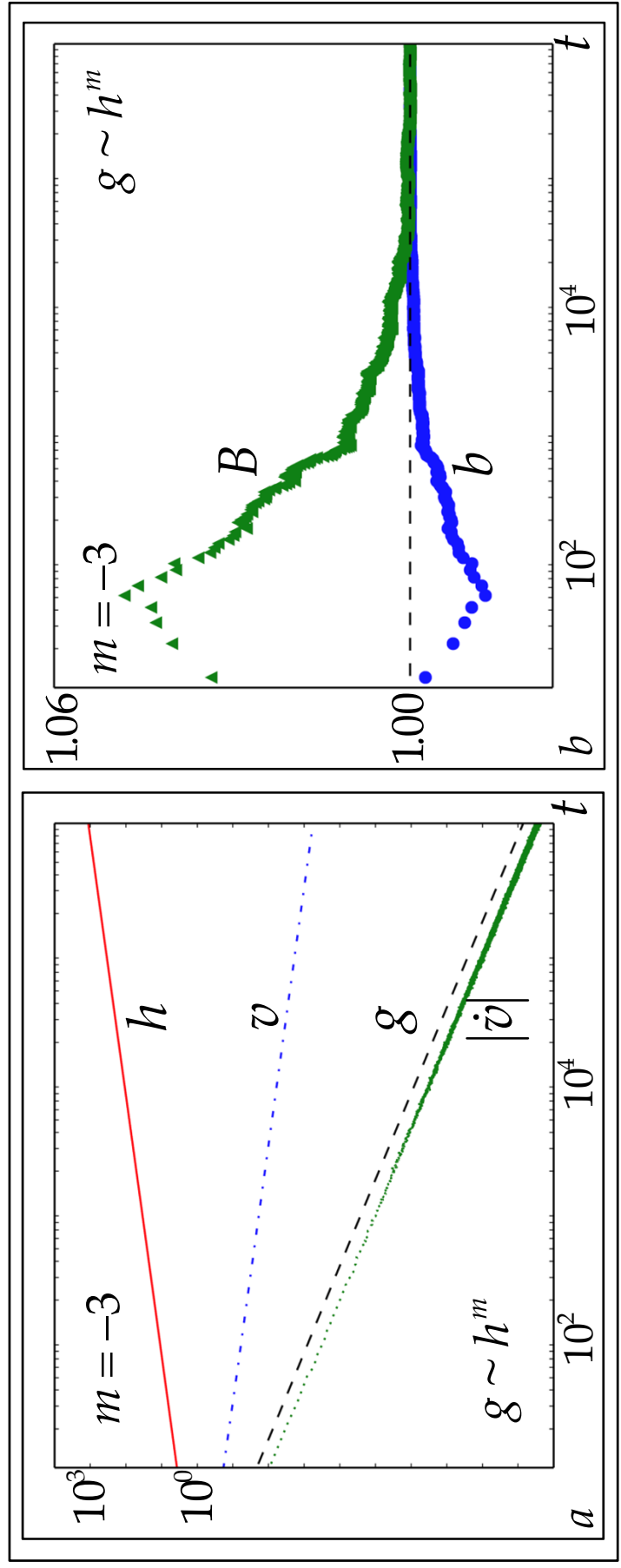

Figure 2

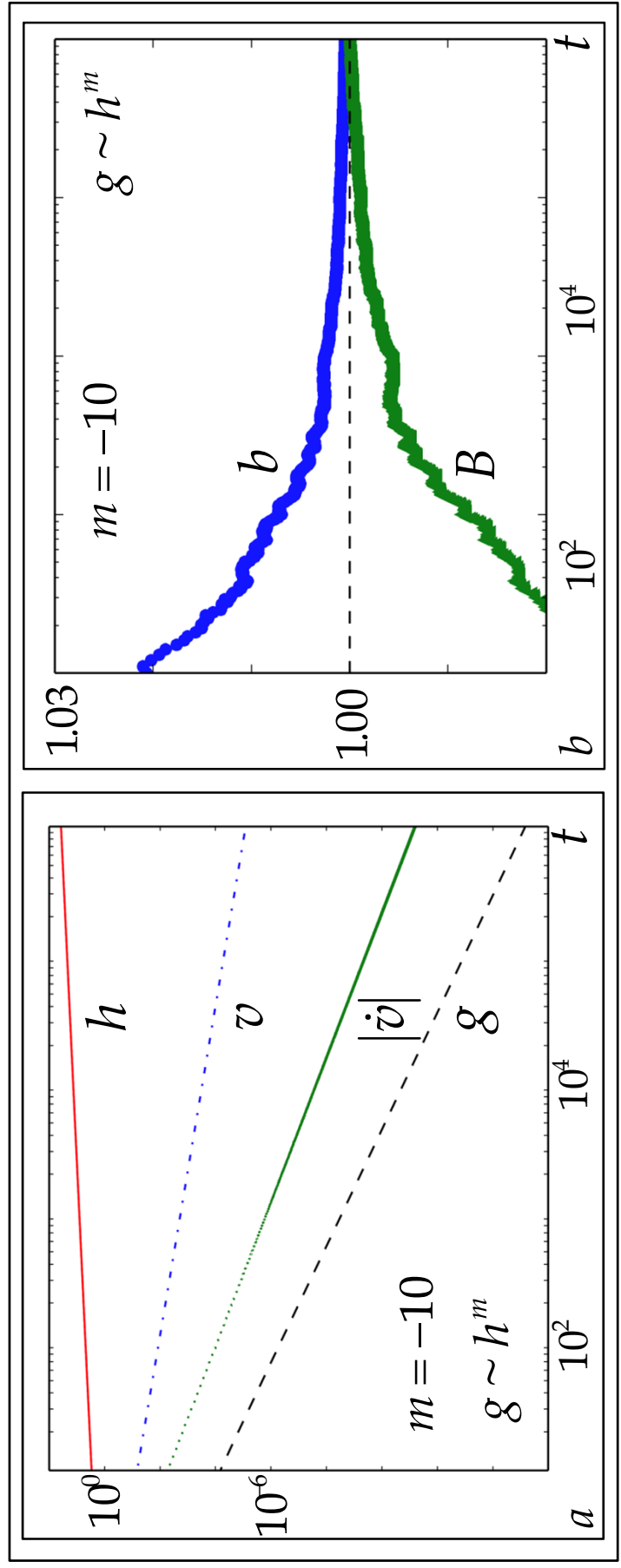

Figure 3

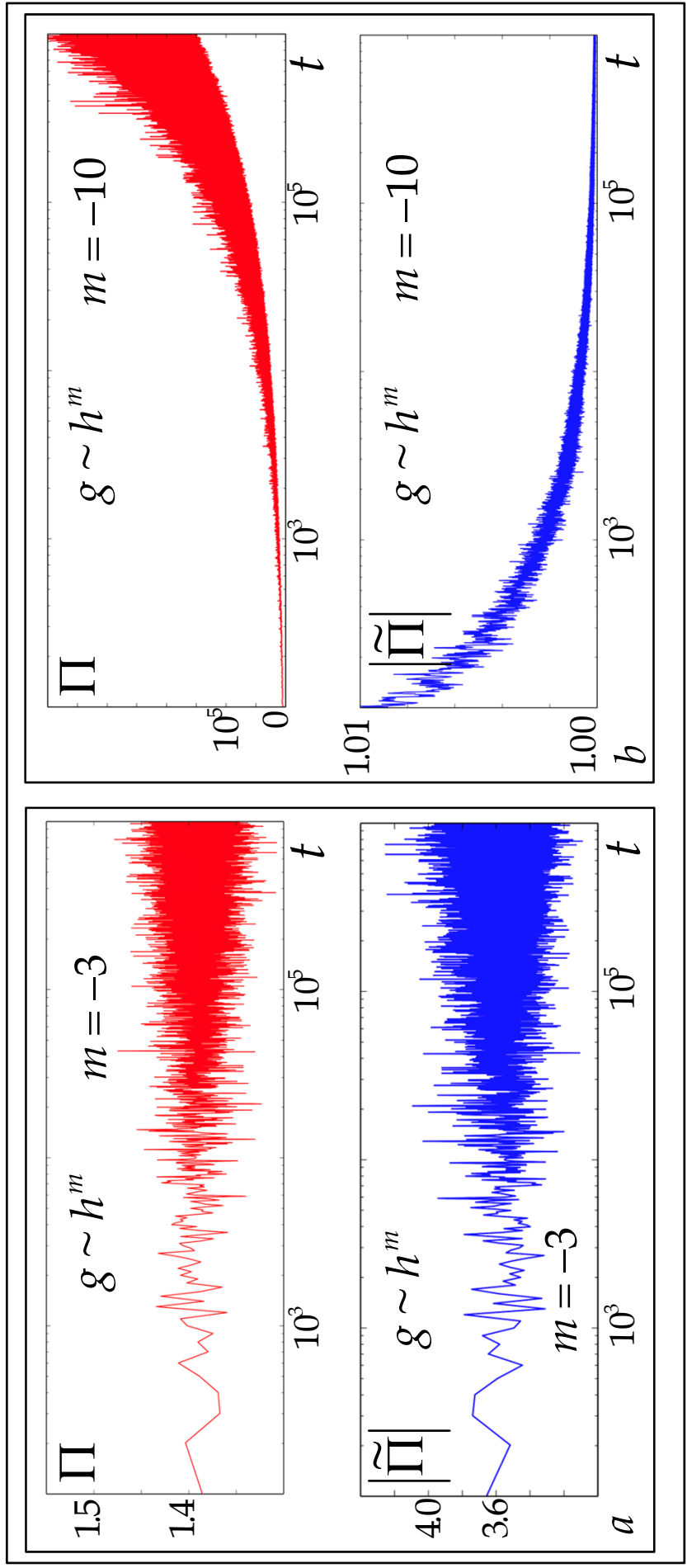